# The solar eclipse of August 30, 1905 at Alcala de Chisvert

Emmanuel Davoust
*Observatoire Midi-Pyrénées*

**Abstract.** Solar eclipses play an important role 1n the history of astronomy because their observation allowed astrophysics to expand. The eclipse of August 30th, 1905 is particularly interesting since the line of totality passes through Spain and Algeria, and that it will last 3 minutes and 45 seconds. Consequently, all the French observatories send astronomers to observe it. About a hundred amateur astronomers of the French Astronomical Society also go to various sites, as well as quite a few individuals who are simply curious, like Camille Saint-Saëns. Count Aymar de la Baume Pluvinel and his two young assistants settle in at Alcala de Chisvert, between Tarragona and Valencia.

**Résumé.** Les éclipses de Soleil jouent un rôle important dans l'histoire de l'astronomie, car leur observation a permis à l'astrophysique de prendre son essor. Celle du 30 août 1905 a un attrait particulier, parce que la ligne de totalité passe en Espagne et en Algérie, et qu'elle doit durer 3 minutes 45. C'est pourquoi tous les observatoires français envoient des astronomes pour l'observer. Une centaine d'astronomes amateurs de la Société Astronomique de France se déplacent également, en ordre dispersé, ainsi que bon nombre de simples curieux, comme Camille Saint-Saëns. Le comte Aymar de la Baume Pluvinel et ses deux jeunes assistants s'installent à Alcala de Chisvert, entre Tarragone et Valence.

When he sees clouds on the horizon in the early morning of August 30th, 1905, Count de la Baume feels discouraged. Will these clouds ruin all his hopes and make all his preparations for observing the eclipse worthless? Should he have gone to Algeria or Tunisia instead?

However, in the month that he has been at Alcala, there haven't been many clouds, except during the second week, and the 16th was the only stormy day. It is true that, the day before, the sky had been cloudy most of the time, but the evening wind restored his hopes. He finally convinces himself that these are the kind of clouds that will disappear after sunrise. And even if the clouds stay, the cooling produced by the eclipse may cause a bright spell at the moment of totality. Twenty-three years of astronomy have taught the Count to be patient. This is his eighth expedition to observe an eclipse; if it fails, he will at least have improved his logistics. And there will be other eclipses, where his participation will be only up to himself.

Indeed, it should be made clear that Count Aymar de la Baume Pluvinel is an independent astronomer; he is not officially attached to any Observatory, and, since 1882, he has spent most of his time and money doing research on photography, the Sun, planets and cornets, simply because it fascinates him.

But he is not an amateur. The Count acquired a solid scientific education at Meudon Observatory, where in 1883 Jules Janssen accepted him as an assistant. His scientific work will be rewarded by his election to the Academy of Science in 1932. From 1913 to 1919, he will be president of the French Astronomical Society.

Choosing the site for the observation of the eclipse was facilitated by an article by the Spanish astronomer, José Landerer, in the Bulletin of the French Astronomical Society, in which the advantages of the different Spanish sites in the zone of totality were clearly presented. Count de la Baume is well acquainted with Landerer because they observed the solar eclipse of 1900 together. The Count trusts him and chooses Alcala, which is easily accessible by train. Jules Janssen chooses Alcosebre, on the coast about ten kilometers from Alcala.

**A prospecting trip**

Preparations begin seriously in April 1905 when Count de la Baume takes a week-long prospecting trip to Spain. In Valencia, José Landerer gives him all the useful information about renting houses and furniture ("don't get cheated about the price", he records in his notebook), about Alcala's magnetic declination and geographical position, on building pillars for the instruments. Landerer also gives him excellent letters of recommendation that will ease things for him in dealing with the key figures of Alcala.

The Count is given a very pleasant welcome at the monastery in Alcala. He is shown the garden where he can install his instruments, the school where he can put boxes, set up the library, clocks and darkroom. The garden is surrounded by a wall which, he hopes, will shelter the astronomers from curious onlookers, whom the Count finds intrusive. "The slightest actions of the astronomers interest them more than the motion of the stars; furthermore, these nosey people believe that astronomers have some kind of understanding with the sky and that, near them, they will see the phenomenon better."

The rustic cells of the Franciscan monastery, with a saint's name above each door, will be suitable for his helpers and the astronomers from Nice. The Count and the heads of the two other missions, Martial Simonin (Nice Observatory) and Nicholas Donich (Pulkovo Observatory) will stay in the villa of a wealthy Spanish family.

The food problem remains unsolved. The Count mentions it to Landerer: "If we ask the monks, I'm afraid that we would be very badly served, and not to our taste. I believe that it is very important to have good food, because we will have to work very hard, and one works well when one is well fed."

He also goes to Alcosebre to visit possible lodgings for Janssen. The trip in a tartan along paths full of holes is extremely uncomfortable. On the way back to France, the Count visits the brand new Fabra Observatory in Barcelona, which had been inaugurated by the king just one year earlier.

**The voyage**

De la Baume brings a whole collection of instruments to observe the eclipse: two coelostats for spectroscopy, a theodolite to precisely measure Alcala's coordinates and to set the orientation of the pillars, an equatorial with a focal length of 12 meters to photograph the solar corona along with a camera for photographic plates measuring 40x40 cm. He also brings tent material, tools, plates and plate holders, equipment to treat and develop the plates, a phonograph to use as a talking clock during the eclipse, books, clothes - all in about twenty crates.

On July 30, 1905, after spending a week carefully packing all this equipment, the astronomers leave Paris by train with "half-price" tickets obtained for them by the French Astronomical Society. Count de la Baume takes a Sleeping car via Avignon with the Russian astronomer, Donich. His two assistants, Fernand Baldet (age 20) and Albert Senouque (23), as well as his valet, Baptiste Longuet, travel sitting up in second class. They go through Toulouse, where they miss the express train for Narbonne.

Count de la Baume describes the rest of the voyage in his cursory way.

*July 31*. Lunch at Cette [Sète] where we find the astronomers from Nice. At Narbonne where we were to find Senouque, Baldet et Baptiste Longuet, there is nobody. At Port-Bou, no parcels. With

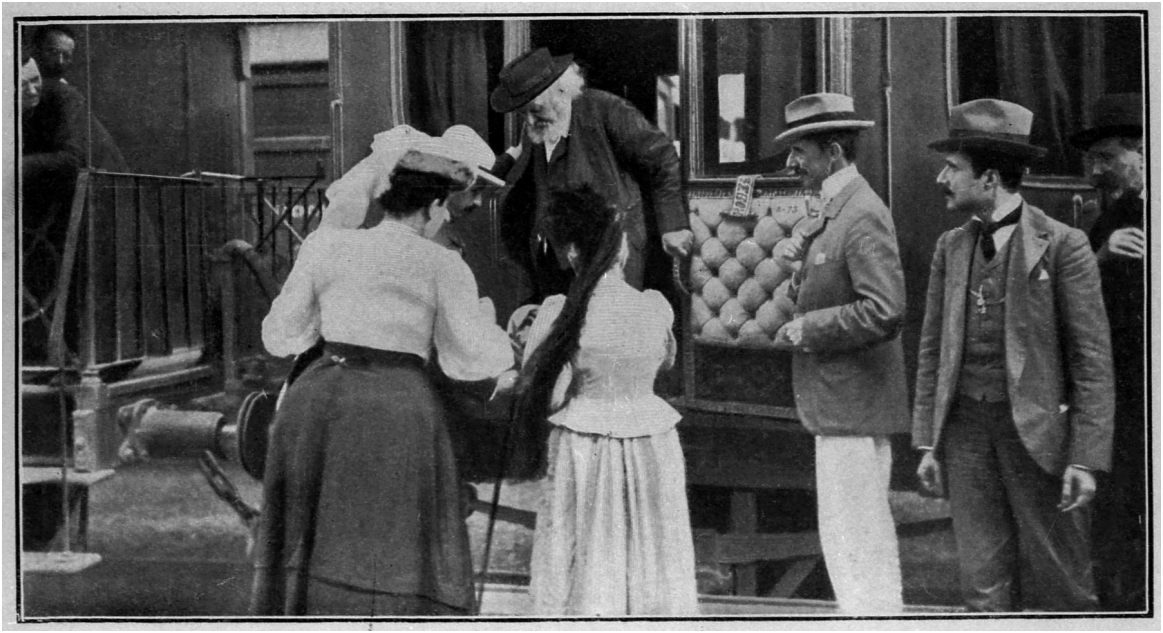

*Count de la Baume Pluvinel welcomes Jules Janssen at the train station on August 18, 1905*

Simonin, we walk to Cerbère across the mountain. There is no road suitable for vehicles from Port-Bou to Cerbère. We learn that the parcels just left for Port-Bou. Return on foot to Port-Bou where Senouque, etc. arrive. Cloudy weather all day. Sleep in horrible hotel. Donich continued on to Barcelona.

*August 1*. At 8 am. we deal with the baggage. We wait for the customs agents. The three of us and Baptiste leave at 12:40 for Barcelona where we arrive at 8 o'clock. All night, while traveling, we see lightning on the horizon. But the sky is almost completely clear.

*August 2*. Arrival at 4:40 am. at Alcala.

**A month in Alcala**

The three missions - the Count's, that of Nice Observatory (Martial Simonin, Pierre Colomas and Stéphane Javelle) and that of Pulkovo Observatory (Marquis Donich and his assistant, Baron von Pahlen) - all share the monastery garden, which measures 45X100 m.

The astronomers start working on the very first day: determining the meridian by observing the Sun, confirming the Sun's passage at the meridian with the chronometer from Nice, measuring the magnetic declination, determining the location of the pillars to hold the telescopes.

On the second day, masons begin to build the pillars with bricks at six o'clock in the morning. In the evening, the instruments have still not arrived. The Count is about to leave for Tarragona when a telegram arrives saying that the instruments will be there the next day. The next morning is devoted to bringing the crates belonging to the Count and to Nice observatory from the station.

The installation is studded with small incidents that the Count notes. "Saturday, August 5. Senouque who did not feel well yesterday has stuck a nail in his foot and does nothing all day. Baldet unpacks. I order the installation of the tents, the Gautier coelostat and the equatorial. The unpacking is finished in the evening. Clouds morning and evening, very nice at one o'clock. A lot of wind in the evening and at night. We have to tie down the tents. The blue tent collapses." The Count also carefully notes the condition of the sky several times a day. All his thoughts are concentrated on the event of August 30th. "Sunday, August 6. A walk in the direction from which the eclipse will come." He does not take advantage of his stay; at the beginning, despite a French cook brought from Barcelona, the food makes him ill.

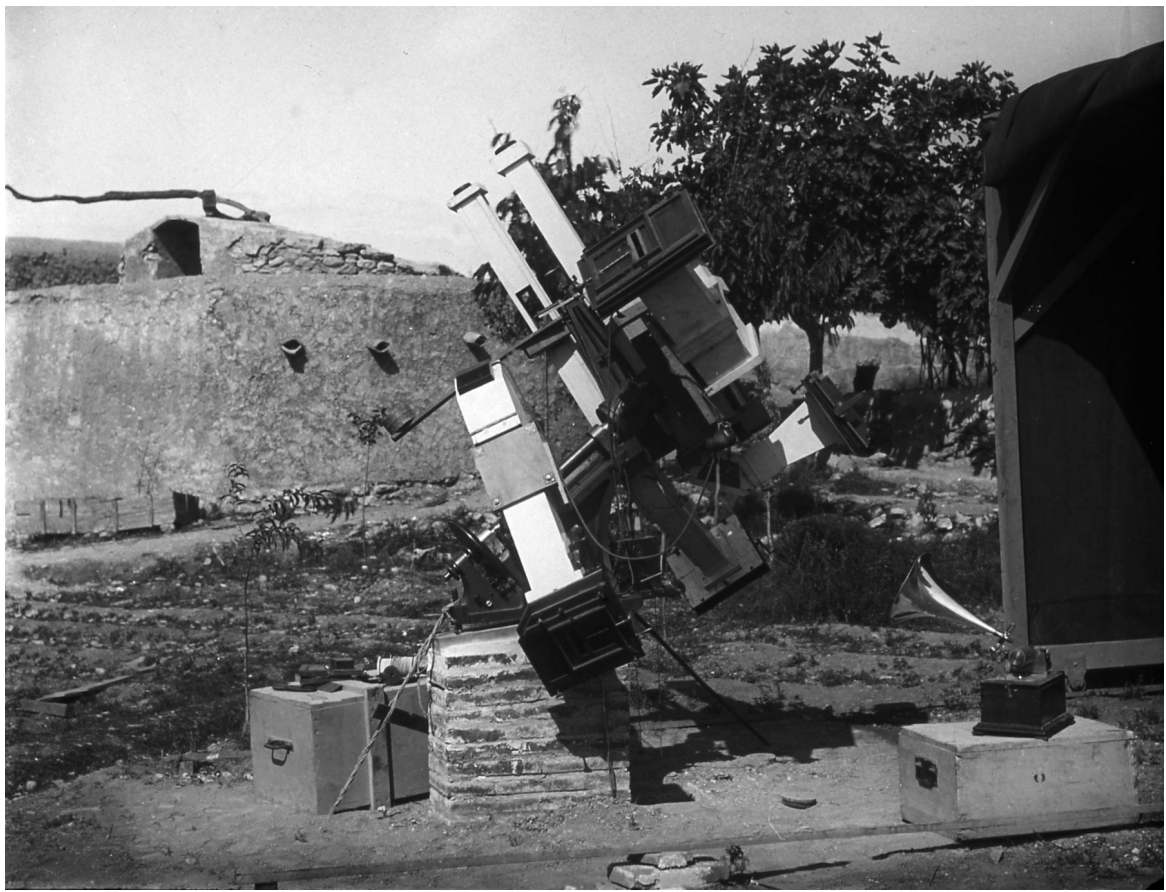

*The instruments of count de la Baume and the phonograph*

The astronomers become acquainted with the town. 6000 inhabitants. Whitewashed houses. A maze of streets. Squares with a fountain in the middle, surrounded by women and children getting water. A special smell in the air; the bakers heat their ovens with fragrant herbs. A cool breeze coming from the sea.

One Sunday, the whole mission goes bathing at Alcosebre, by tartan. All the pleasure is for Fernand Baldet, who describes the vehicle. "Imagine the wagons of our peasants, but smaller and much more original. Inside there is a padded bench on each side, no floor but matting loosely held by ropes all around. Above, an arched roof, most often made of bamboo. There are no springs whatsoever."

On August 7th, the instruments are in place, and from that date on, observations take place every evening of good weather until midnight. The instruments have to be placed in station, that is, the axis of rotation must be set precisely according to that of the Earth, and then they must be balanced so that the tracking is correct. Next, the various mirrors of the coelostats must be adjusted, the prisms put in place and the spectrographs be set. They choose to observe bright stars with the same declination as the Sun, Andromeda, Altair, in order to be as close as possible to the observational conditions of the eclipse. By tracing the trajectory of these stars on some opaque glass, the orientation of the plates with respect to the North can be determined. Count de la Baume, who is meticulous and demanding, leaves nothing to chance.

On August 18th, Jules Janssen, director of Meudon Observatory, arrives with his wife and daughter; he is 81 years old. Raphaël Bischoffscheim, also in his eighties, arrives on the 28th. He is a banker, a patron of French astronomy, the founder of Nice Observatory. His is accompanied by Henri Perrotin, the son of the former director of Nice Observatory, and by Mr. Huret, a journalist of the Figaro newspaper, and his wife. On the other hand, General Bassot, director of Nice Observatory, who is also expected, will not come.

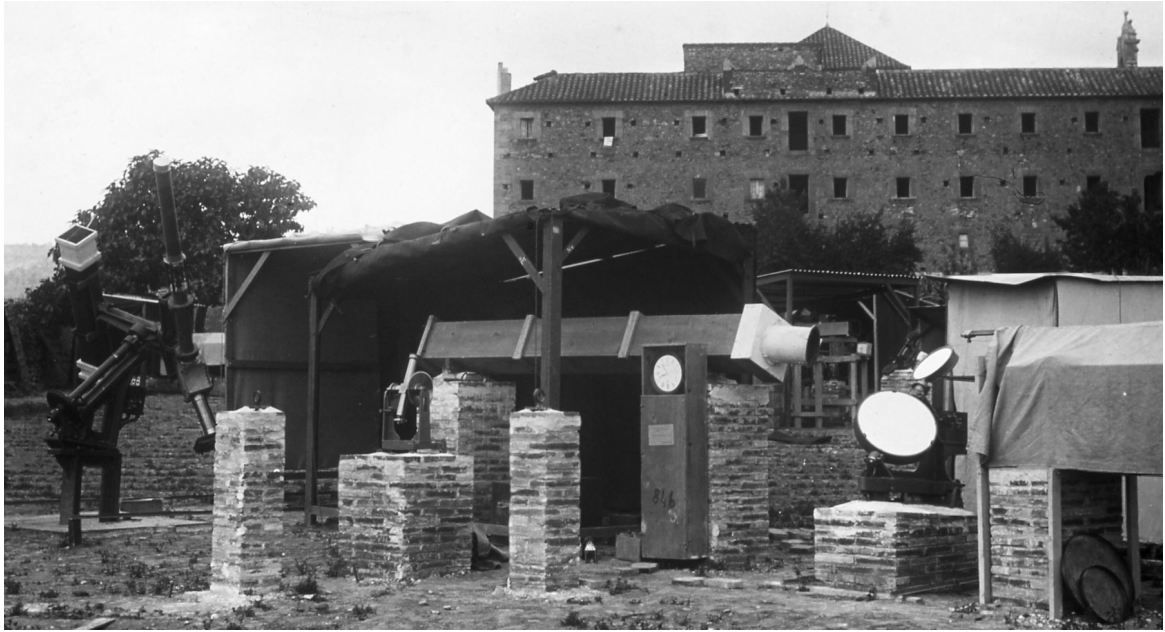

*The instruments of Nice Observatory*

Preparations continue. On August 20th, test photographs are made using five different brands of plates. Jougla plates turn out to be the best; Simonin sends a telegram to Bassot to get more of them.

During the final week, silvering baths are prepared, then the mirrors are silvered, the prisms are set, the slits in the spectrographs are positioned, the orientations of the plates in the cameras are determined.

The day before the 30th, the plates are sensitized, the glass side is covered with an antihalo, and they are placed in the plate holders. There are rehearsals, so that gestures are automatic during the eclipse to avoid errors. It is decided not to use the phonograph which still does not want to work, despite efforts by Fernand Baldet. He does, however, manage to repair Donich's recording barometer, who promises him the gratitude of the Imperial Academy.

On the morning of the eclipse, the lenses are cleaned with a shaving brush, the clock movements are wound up. Count de la Baume and Fernand Baldet tell us the rest of the story.

**The eclipse seen by Count de la Baume**

"We are still loading the plate holders. (We should have started earlier.)

1/4 of an hour before the eclipse, clouds appear in the West and become menacing. I observe in the visual spectroscope and in the refractor. (.) I keep looking at my watch to know when the eclipse is going to begin. We should have had somebody to count the remaining minutes and seconds. I take an exposure on the 6pr for a comparison ten minutes before totality. A 5 second exposure. Much too long.

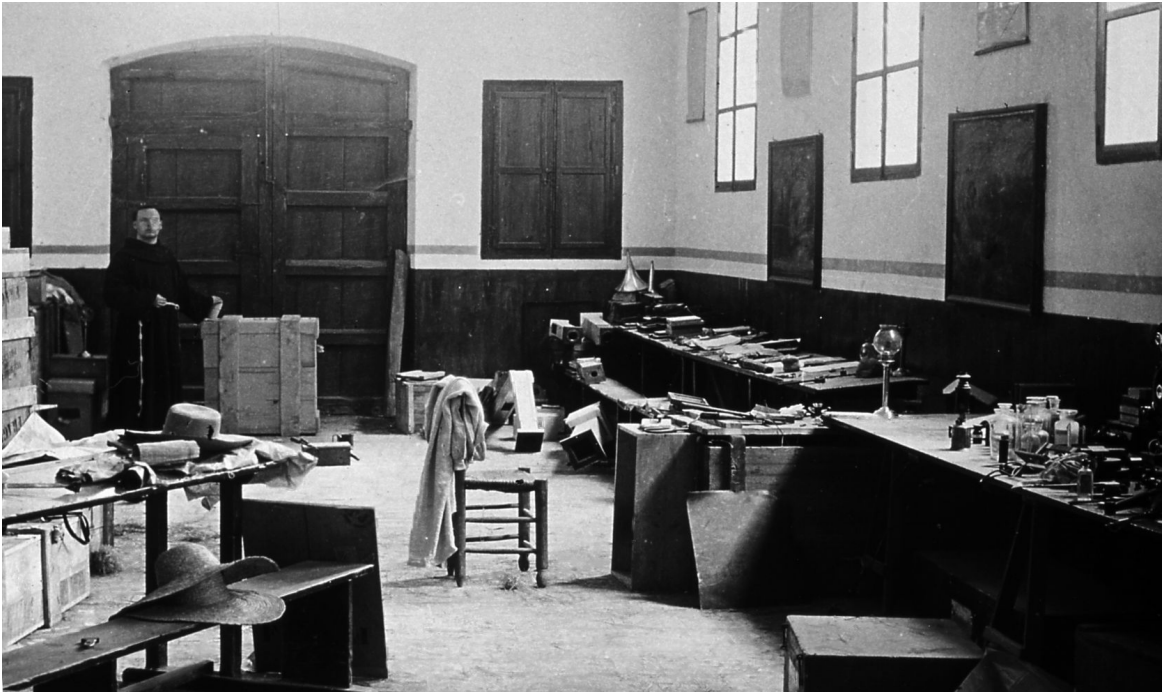

*The classroom where the instruments and straw hats of the astronomers are stored*

Javelle shouts, "Minus ten seconds" about 30 or 40 seconds before totality. He was supposed to call out when the cord of the indentation was 23 mm (Simonin's calculation should be redone). I see right away that there are more than 10 seconds before totality and call to Colomas, "Not so fast", so that he won't use up all his plates before totality (which he did). I see several Bailey grains and 1 shout "Go" when the last grain has disappeared. Very busy looking in the telescope, I did not look into the spectroscope. Immediately, I freed the hand on my watch, forcing me to look at it (great loss of time). I press the button on the Flint, I open the Cooke, I open the 6pr, I go back to close and change the Cooke, I close and load the Flint[1]. * Then I go out to look and am disappointed to see a cloud on the Sun. 1 continue with the operations. At 3:20, 1 go to close the Flint and since I see the light returning fast, I run to close the 6pr and the Cooke. Then I come back to the phosphorescence. At 3:30, the Sun has completely reappeared."

**The eclipse seen by Fernand Baldet**

Fernand Baldet is to take photographs at the equatorial with a focal length of 12 meters. Since morning, he has been practicing placing and removing the photographic plate holders on the equatorial, along with a young monk who will assist him during the eclipse.

A letter to his parents tells us how he lived through the event.

"At 11:55, the moon begins to touch the Sun. At 12:35, more than half the Sun is hidden. I realize that it is darker.

Now the light gets weaker and weaker.

Half an hour before, we go and get the photographic plate holders. Last rehearsal. I no longer have a phonograph, we were not able to record on it very well. In fact, I prefer not to have it. I have a chronograph and a metronome that beats the seconds.

---

[1] The Cooke, the 6pr and the Flint are spectrographs.

At 1:00, shouts of "Light the lanterns!" We make everybody move away, and the policemen tell the people on the rooftops to be quiet. Now I must not lose my head. We notice clearly that the light is disappearing. The show is becoming impressive. Minutes go by, it is getting darker, everything is gloomy.

1:05, Mr. de la Baume calls, "Three more minutes". Now it is getting dark extremely fast, I am at my position.

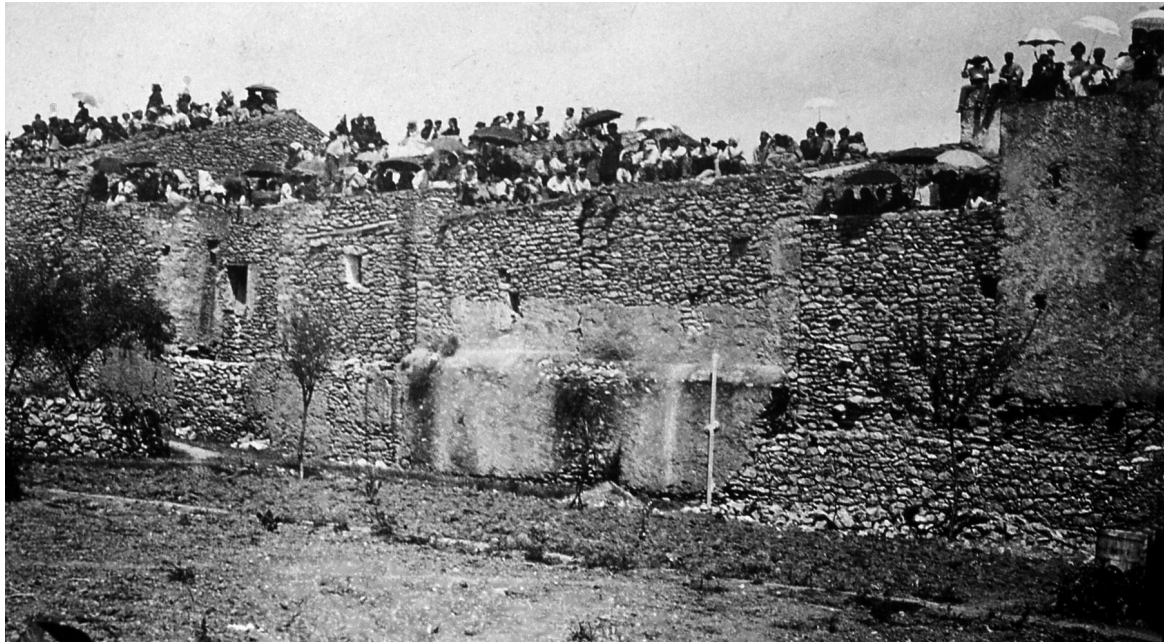

*The crowd on the roofs of Alcala watching the astronomers half an hour before the eclipse*

Mr. Javelle, from Nice Observatory, calls, "Minus ten seconds", the Count is at his telescope to shout out the "Go" of the beginning. The light goes ont like a lamp, there is complete silence, the moment is unforgettable. The sky is pure. I open my plate holder. My hand is on the shutter, the metronome counts the seconds.

All of a sudden, at the same time as the last ray goes out, a scream rips through the air -- "Go!" The Count shouted so loudly that the sound was strangled in his throat. I open, expose for 2 seconds, close, give the plate holder to my assistant who goes to put it away, I take the second one without hesitating, expose for 7 seconds, close, give it to my assistant, take the third one and run to my chronograph that will count a hundred seconds, I open and go out.

1 remain dazed when I see a very light cloud passing in front of the Sun. However it is very visible through it. Instead of a brilliant disc, a perfectly black disc hangs in the sky, surrounded by the solar atmosphere and indented by prominences. The stars shine, the horizon is as red as during a beautiful sunset, and the clouds are purple. The show is gorgeous, one has to see it to understand what I mean.

My hundred seconds go by slowly. At the 90th, I get ready to close, and do so at the hundredth. I take my fourth plate holder, 40 second exposure. I go out again, the cloud has neither decreased nor increased. Finally the fifth exposure for 11 seconds. I did everything mathematically, I made no mistakes. I go out, I should have 18 seconds to myself before the end, I look, the first ray appears almost as soon as I go out. Gorgeous show. The light increases rapidly. The eclipse lasted 15 seconds less, we can't believe it. Many of the plates will be fogged because of this. The Count asks me if my plate holder was closed, if I had had enough time, fortunately yes.

Five minutes later, the sky is completely cloudy, and one hour later, for the first time, if we don't count the storm on August 16th, a gentle rain is falling. We go for lunch."

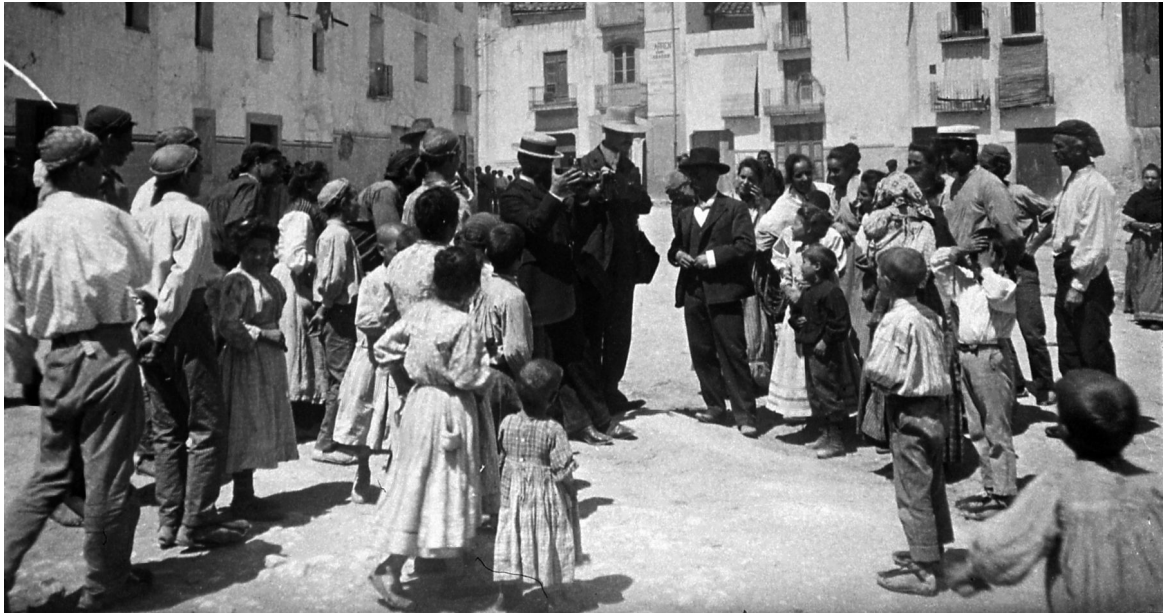

*The French astronomers surrounded by a crowd of curious people*

**Epilogue**

The following days, astronomers pack the instruments and develop the plates. The members of the expedition leave progressively : von Pahlen on September 4, Donitch on September 5, the count de la Baume and his team on September 6. Emile Senouque travels directly to Janssen Observatory on Mont Blanc, to study the variations of the magnetic field of the earth. Fernand Baldet returns to Paris where he is the apprendice of a jeweler. He will analyse the observations the following year in the laboratory of count de la Baume Pluvinel.

**Acknowledgments**. This is the translation of an article in French published in *L'Astronomie*, 1995, 109, 309-313. We thank Jeanne Lagarde for providing her father's notes, letters and photographs relative to this expedition.